# Pulse-induced switches in a Josephson tunnel stacked device

G.P. Pepe, G. Peluso, M. Valentino and A. Barone

*I.N.F.M.-Dipartimento Scienze Fisiche, Università di Napoli Federico II, Naples, Italy*

L. Parlato

*Seconda Università degli Studi di Napoli, Facoltà di Ingegneria, Aversa ( CE ), Italy*

E. Esposito, C. Granata and M. Russo

*Istituto di Cibernetica del CNR, Via Toiano, Arco Felice, Naples, Italy*

C. De Leo and G. Rotoli *

*Dipartimento di Energetica and \*I.N.F.M. , Università dell'Aquila, Località Monteluco, Roio Poggio, I67040 L'Aquila, Italy*

*Abstract*

Pulse activated transitions from the metastable to the running state and viceversa have been observed in a stacked double tunnel Nb-based Josephson system. Experimental results are compared with numerical simulations based on the Sine-Gordon model of the stacked junctions by injecting pulses with variable amplitude in one of the junctions of the stack, and observing the voltage response of the other junction. Both experimental and numerical results show the possibility to induce both direct and back-switching transitions from the metastable to the running state simply by changing the amplitude of the electronic pulses injected across the stack device.

A Josephson junction biased by a fixed external current can be considered as a prototype of a macroscopic multi-stable system. Besides the effects due to fluctuations[1], widely studied in the past, and the fundamental tests of quantum mechanics[2], the study of the behavior of the escape from the zero-voltage state could deserve new exciting aspects for basic physics not only in view of device applications but also for novel physical effects involving non-equilibrium dynamical states. Double tunnel junction configurations have been widely used both for non-equilibrium fundamental experiments[3] and several transistor-like superconducting devices.[4-5] They offer the possibility of changing directly the electronic distribution function of the intermediate electrode by combining the quasi-particle injection as a non-equilibrium source, the phonon relaxation and the following Cooper pair breaking.

In this note we present both experimental and numerical results concerning the behavior of a current biased Josephson Junction (JJ) under the influence of an electronic pulse injected from a second JJ forming a double tunnel junction stacked device. From a non-equilibrium point-of-view, one junction is used to break Cooper pairs through the injection of quasi-particles into the middle common film; the second junction, independently biased at a value lower than the critical current, senses the effects due to a change in the electron distribution function, switching thereby to the finite voltage state. We have studied the passage from the Josephson (V=0) to the dissipative state produced by injected electronic pulses, and also the possibility these device offers of resetting under suitable conditions for injected electronic pulses in terms of their amplitudes. The experimental results have been checked by a numerical approach based on the Perturbed Sine-Gordon Equation (PSGE) for each junction in the stack, where the influence of the electronic perturbation has been taken into account simply by pumping one junction in the stack with a train of variable amplitude pulses and then observing the effect on the other junction due to the stack coupling. To our knowledge it is the first time that a pulsed configuration is used in a stacked system to investigate the properties of the Josephson current and its lifetime in a current biased junction mode.

Stacked tunnel JJs have been fabricated by following a novel process developed for high-quality three terminals superconducting electronics[6]. This process will be now briefly outlined. The substrate used was a 2inch crystalline Si wafer held at room temperature during films' deposition. The whole penta-layer structure, i.e. Nb-AlxOy-Nb/Al-AlxOy-Nb, was deposited *in situ* in a UHV system whose backpressure was $P=1.0 \times 10^{-8}$ Torr, without breaking the vacuum. Nb and Al films were deposited by magnetron sputtering, while tunneling barriers were formed by thermal oxidation of the Al films. The final structure consisted of Nb(150nm)-Al(10nm)-AlxOy-Nb(40nm)/Al(20nm)-AlxOy-Nb(40nm) After the penta-layer deposition, the photoresist was removed with a lift-off process, and hence the geometry of the bottom electrode was obtained. By dry and chemical etching processes, the top Nb film, the thin AlxOy oxide and the Al film were removed in sequence from the top with the geometry of the top JJ. An anodization oxide produced the isolation of the bottom junction and also its area, with the exception of a small hole for contacting the intermediate electrode. Afterward, SiO bridges were deposited by thermal evaporation in order to have a sort of planarization and a further isolation in the next wiring process. The wiring of both the intermediate and the top electrodes was obtained, after a sputter cleaning process, by depositing a layer of Nb (350nm) photo-litographically defined by a lift-off process. Finally, the intermediate Nb film and the bottom oxide barrier outside the junction area were removed by a R.I.E. etching process. The final cross section of the realized stacked junction is shown in Fig. 1.

The samples were characterized in a liquid Helium cryostat with a local μ-metal shielding. The main results are summarized in Table I. The ratios between the static tunnel resistance at V=1mV, $R_S$, and the normal tunneling resistance, $R_{NN}$, are reported together with sum-gap voltage, $V_g$, its width $\Delta V_g$, the measured current jump at $V_g$,(see $\Delta I_g$) and the critical current $I_c$. All data were recorded at T=4.2K.

The Josephson penetration depth $\lambda_J$ was estimated to be 70 μm, thus implying a ratio $L/\lambda_J \approx 1.4$ for a Josephson critical current $J_c = 80$ A/cm$^2$ and L=100 μm. The junction capacitance, estimated from the position of the first Fiske step ($V_1$=45μV), was about nF [7].

The bottom junction was used as injector of current pulses, whose effect was observed across the top junction (detector). The injected pulses had rise-times of 2 ns, 600 μs of duration, delay time 20 ms. They were supplied by a conventional pulse generator (EG&G Mod. 480). Each pulse was split and sent simultaneously to both the injector junction and to a digital oscilloscope (Le Croy Mod. 9361, 300MH) for triggering the waveforms' acquisition. With the detector biased at I<Ic, the output voltage across the detector was measured by a standard PAR Mod. 5113 pre-amplifier (300 kHz BW). In Fig. 2 we report the output detector voltage for different injection pulse amplitudes. Apart some reflections in the signal due to a not completely matched injection line, a clear correlation between injected pulses and the detector voltage is present. In particular, in the presence of injected pulses across the bottom junction, a change of the detector voltage from the state V=0 to the quasi-particle branch of the I-V curve and *viceversa* is observed (see Fig. 2 a-c). This behavior, observed both in single shot and a.c. steady-state measurements (-the delay time between pulses was $T_{delay}$ = 20 ms-), cannot be ascribed to the specific loading of the detector junction (-the load line remains the same during the switching measurements-). For larger amplitudes no back-switching is observed, and the detector junction still remains on the dissipative state at any pulse across the injector.

For amplitudes within a suitable range the device has a *flip-flop*-type logic characteristic, i.e. it commutes for each pulse across the injector. The possibility to control the voltage state of a JJ by pulse injection through a second junction can have many potential applications ranging from nuclear integrated superconductive detectors[8] to several logic devices.

In order to model the double tunnel junction system we describe it as a stack of two long junctions using the theory developed in recent years for stacked junctions[9-11]. In fact, inductive couplings between two or more junctions in stacked configurations have been observed and explained in long Josephson junctions. Coupling between stacked junctions depends on the second spatial derivative of the phase. This means that the single analog pendulum for small junction is hardly a correct description. In the following we model the junctions in the stack as long junctions. This allows spatial variation of the phases, and so of magnetic fields and currents. For sake of simplicity we assume that injection and bias currents have an equal well-defined direction so the spatial variations take place only in one of the spatial dimension of the junction avoiding the use of more complex two-dimensional description. In normalized units the PSGE's for injector, $\psi$, and detector, $\varphi$, are:

$$\varepsilon\partial_{xx}\varphi - \partial_{xx}\psi + \partial_{tt}\psi + \alpha\partial_t\psi + \sin\psi + \gamma_P(x,t) + \gamma_N(x,t) = 0$$
$$\varepsilon\partial_{xx}\psi - \partial_{xx}\varphi + \partial_{tt}\varphi + \alpha\partial_t\varphi + \sin\varphi + \gamma_B + \gamma_N(x,t) = 0$$

where $\varepsilon$ is the stack coupling, $\alpha$ the loss parameter, $\gamma_B$ the constant detector bias. We note that the injector junction is not biased, but the function $\gamma_P(x,t)$ represents the effect of pulses. The function $\gamma_P(x,t)$ is chosen linear in space and in the form of a triangular pulse train in time, i.e.

$$\gamma_P(x,t) = \begin{cases} 2\Gamma\left(\dfrac{x}{l}\right)\left(\dfrac{t}{T}\right) & \text{for } 0 \leq t \leq \dfrac{T}{2} \\ 2\Gamma\left(\dfrac{x}{l}\right)\left(1-\dfrac{t}{T}\right) & \text{for } \dfrac{T}{2} \leq t \leq T \\ 0 & \text{for } T \leq t \leq T_p \end{cases}$$

where $T$ is the pulse length, $l$ the normalized junction length and $T_p$ the time interval between pulses. This choice implies that pulse is localized mainly on one side of the stack. Peak values $\Gamma$ is the maximum values of current pulse in the injector junction (average values is 1/4 of this). We choose $l=1$ and $\alpha=0.25$ for both junctions (no variation of losses was introduced, even if these cannot be excluded). The term $\gamma_N(x,t)$ is a Gaussian noise at 4.2 K modeled as in Ref.12. By symmetry this was added to both junctions also if it has no effect on the unbiased junction. We take $\varepsilon=-0.85$ for the stack coupling, and no field at the junction edges (open boundary). The detector is biased at $\gamma=0.2$, which is sufficient to exclude any effect of thermal return current.

Results are reported in Fig.3: from a) to c) the plots show the spatial average voltages in both junctions. A train of pulses of normalized time length $T$ equal to 100 was applied in the injector junction with peak amplitudes slightly increasing from $\Gamma$ equal to 20.4 (a) to 21.0 (c). The voltage pulse response of the injector junction is shown as full squares line in Fig.3. The pulse response of the detector junction is shown with full circles. The flip-flop state progressively sets on with the increase of the pulse amplitude in agreement with the experiments. We note that similar results can be obtained indifferently also with different values of the stack coupling $\varepsilon$ even if at different pulse amplitudes and/or pulse spatial dependence (e.g. using a quadratic rather than linear pulse, has no qualitative effect). Change of the pulse length has no effect since the detector junction always responds on the pulse trailing edge.

Simulations show that a moderate increase in temperature has the effect of smearing the set on of flip-flop state in amplitude. Moreover the same transition from no switches to fully developed flip-flop state is obtained also sweeping the bias current at fixed pulse amplitude. These properties are very reminiscent of thermal escape from Josephson current in small or long junctions[12-13]. So an interesting hypothesis can be stated in the case of direct switches from Josephson to resistive state: pulses induce in the detector junction a pulse-assisted escape from washboard potential. Pulses increase the energy of the Josephson oscillations causing the escape toward resistive state. As in the normal thermal escape the transition is smeared by the noise. The same idea can be applied also to the back-switching transition. We think that the return current phenomenon[14] is important in order to determine the reset to the zero voltage state, but the influence of heating and self induced fields make the analysis much more difficult.

In conclusion, we presented measurements of pulse activated transitions from the metastable to the running state and *viceversa* in a stacked double tunnel Nb-based Josephson system.. The results have been compared with numerical simulations of PSGE model applied to a stacked system in which the pulse pumping has been modeled as an extra current source with suitable time characteristics. Besides the new interesting physical aspects concerning with the study of both direct and back-switching pulse activated transitions, the device has great potentialities for the development of *flip-flop*-type logic devices, and front-end electronics for nuclear integrated Josephson detectors.

We are grateful to G.Filatrella for useful discussions on the effects of the thermal noise. This work is supported by the I-MURST COFIN2000 Program *Dynamics and Thermodynamics of vortex structures in supeconducting tunneling* .

**FIGURE CAPTIONS :**

Fig. 1: Cross section of a stacked tunnel device with the indication of different layers and electrical contacts.

Fig. 2: The detector output voltage as a function of injected electronic pulses across the coupled junction. The scale is referred to the detector output voltage, while pulses are reported in arbitrary units.

Fig. 3: Simulated flip-flop transition of two junction stack vs the pulse amplitude: (a) $\Gamma=20.4$; (b) $\Gamma=20.6$; (c) $\Gamma=21.0$. The square dot light gray curves refers to injector junction, the circle dot black curves to detector junction. Time and voltage are normalized to $1/\omega_J$ and $\omega_J \Phi_0$ respectively

TABLE :

Table I: Relevant I-V data

|  | Top JJ | Bottom JJ |
|---|---|---|
| Area [$\mu m^2$] | 100x100 | 108x125 |
| $V_g$ [mV] | 2.54 | 2.70 |
| $\Delta V_g$ [mV] | 0.36 | 0.20 |
| $R_s$(@1mV)/$R_{NN}$ , T=4.2K | 20.0 | 36.8 |
| $\Delta I_g$ [mA] | 11.8 | 13 |
| $I_\chi$ [mA] | 7.0 | 9.0 |

G.P.Pepe et al.

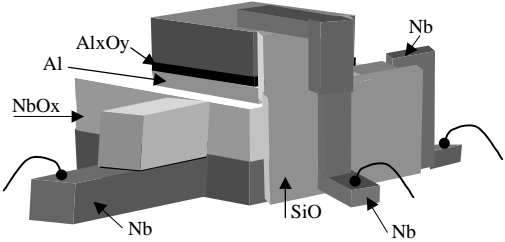

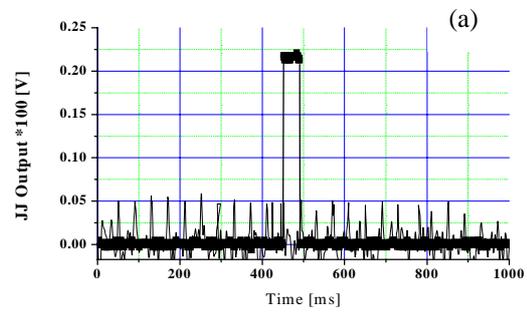

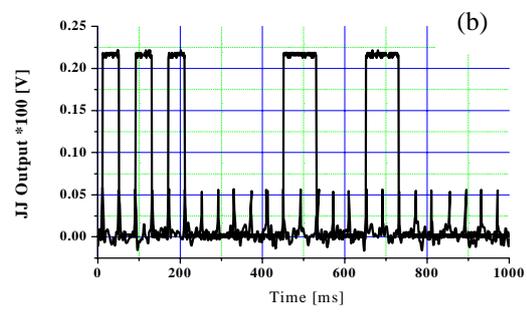

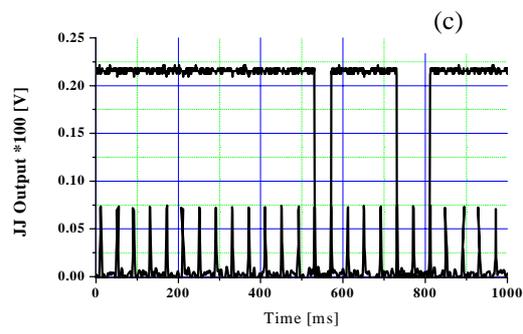

G.P.Pepe et al.G.P.Pepe et al.

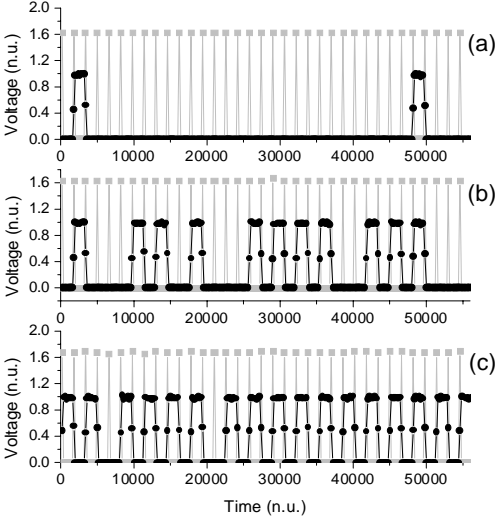